Robert E. Kent

# The Institutional Approach

(in memory of Joseph Goguen)

*Systems, scientific and philosophic, come and go. Each method of limited understanding is at length exhausted. In its prime each system is a triumphant success: in its decay it is an obstructive nuisance.*

<div align="right">Alfred North Whitehead, *Adventures of Ideas*</div>

This chapter discusses the institutional approach for organizing and maintaining ontologies. The theory of institutions was named and initially developed by Joseph Goguen and Rod Burstall. This theory, a metatheory based on category theory, regards ontologies as logical theories or local logics. The theory of institutions uses the category-theoretic ideas of fibrations and indexed categories to develop logical theories. Institutions unite the lattice approach of Formal Concept Analysis of Ganter and Wille with the distributed logic of Information Flow of Barwise and Seligman. The institutional approach incorporates locally the lattice of theories idea of Sowa from the theory of knowledge representation. The Information Flow Framework, which was initiated within the IEEE Standard Upper Ontology project, uses the institutional approach in its applied aspect for the comparison, semantic integration and maintenance of ontologies. This chapter explains the central ideas of the institutional approach to ontologies in a careful and detailed manner.

1.  INTRODUCTION

The institutional approach for the logical semantics of ontologies provides a principled framework for their modular design; in particular, it provides a natural framework for formulating the "lattice of theories" (LOT) approach to ontological organization. According to Sowa, the purpose of LOT is to provide a systematic way of relating all possible ontologies in order to facilitate their inevitable upgrades and conversions. The goal of LOT is to create a framework "which can support an open-ended number of theories (potentially infinite) organized in a lattice together with systematic metalevel techniques for moving from one to another, for testing their adequacy for any given problem, and for mixing, matching, combining, and transforming them to whatever form is appropriate for whatever problem anyone is trying to solve" (Sowa, 2000). The theories of Institutions (Goguen and Burstall, 1992), Information Flow (Barwise and Seligmann, 1997) and Formal Concept Analysis (Ganter and Wille, 1999) have independently formulated and developed various concepts surrounding the LOT construction. But the institutional approach gives the most comprehensive treatment.

Within an institution the lattice of theories is the indexing of the context of theories by the context of languages (an index can be thought of as a list or a pointing device; a context (Goguen, 1991) represents a kind of mathematical structure, such as algebraic structure, topological structure or logical structure; contexts and indexed contexts are discussed below). The central relation inside the lattice of theories is logical entailment. However, in problem solving we are always supposed to be willing to "think outside the box." In this situation, the institutional approach instructs us to think outside the lattice of theories. Inside is the entailment relation between theories, but outside are links between theories. Theory links specialize to theory entailment within the fiber over a language (this fiber consists of all theories having that language; the idea of a fiber is defined in the section on passages). Theory links (discussed further below) are logical language maps between theories that preserve entailment. Theories and links between theories form the context of theories.

The institutional approach starts with the motivation to unify the numerous efforts to use logic for the representation and organization of the knowledge space of various communities. In order to accomplish this, the institutional approach uses the two related distinctions (the general versus the special) and (the abstract versus the concrete). Both generalization and abstraction can have many levels or degrees. The correct level or degree depends on the goal in mind. In order to reach a certain level of generality we need to abstract from the unimportant and superfluous details, but still retain the essential ones. In the institutional approach, a logical system is identified with an institution. The institutional approach, whose goal is the representation and maintenance of systems of ontologies, generalizes by abstraction over various logical systems such as first order, second order, higher order, Horn clause, equational, temporal, modal and infinitary logics.

This chapter discusses the institutional approach within the theory and application of ontologies. One caveat: although the institutional approach to ontologies extensively uses category theory, this chapter has not been written for a reader with background knowledge of category theory. Instead this chapter has been written for philosophers, computer scientists and the general public. For this reason less technical and more common terminology has been used in describing the basic concepts. Such an approach has been used before. Goguen has used the term "(mathematical) context" for the category concept (this may be an especially useful alternative for philosophers), Manes has used the term "passage" for the functor concept, and Lawvere and others regard the concept of adjoint functors as a generalized "inverse pair" of functors. There is a key$^*$ for this terminology in the endnotes. In addition, very few abstract symbols have been used. This is a mixed blessing, since, although the intimidation of abstract mathematics has been removed, the advantage of the extremely useful idea of "commutative diagram" (studded with abstract symbols) cannot be easily used. For example, the general discussion about the architecture only uses the symbol **V** for a general ambient context and the symbol **Cls** for the specific ambient context of classifications (an ambient context is a background encompassing context within which we form diagrams; the institutional approach uses the ambient context **Cls** of classifications and infomorphisms). For an overview of category theory, see the chapter by Healy.

### *1.1. Ontologies*

Ontologies are of two types: populated and unpopulated. Populated ontologies contain instance data, whereas unpopulated ontologies do not. Instances (tokens, particulars) are things that are classified, whereas types (universals) are things that classify. 'Aristotle' is a particular individual in the ancient world, whereas 'human' and 'philosopher' are types that classify and describe that individual. "It is particulars, things in the world, that carry information; the information they carry is in the form of types" (Barwise and Seligman, 1997).

Any ontology is situated within the context of the logical language of a domain (of discourse), which often consists of the generic ideas of the connectives and quantifiers from logic and the specific ideas of the signature (the constant, function and relation symbols) for that context (Goguen and Burstall, 1992). An unpopulated ontology is represented as a theory consisting of a collection of statements or sentences based on the language. The theory allows for the expression of the laws and facts deemed relevant for the domain. A structure of a domain provides a universe of discourse in which to interpret statements of a theory. Both theory and structure are described and constrained by the logical language. A populated ontology is represented as a (local) logic or knowledge base having two components, a theory and a structure that share the same underlying language. This notion of logic is a precursor to the local logics defined and used in Information Flow (Barwise and Seligman, 1997), which are more closely represented by the composite logics defined below. In general, the logics in the institutional approach are neither sound nor complete. A logic is sound when each sentence of the theory is true when interpreted in the structure; that is, when the structure satisfies each sentence of the theory. A logic is complete when every sentence satisfied by the structure is a sentence entailed by the theory. Associated with any structure is a natural logic whose theory consists of all sentences satisfied by the structure. The natural logic is essentially the only sound and complete logic over a given language. Associated with any logic is its restriction — the sound logic with the same underlying structure, whose theory consists of all sentences satisfied by the structure and entailed by the theory.

There is a projective component passage from logics to structures. In the opposite direction, there is a natural logic passage from structures to sound logics. With structure projection and natural logic, the context of structures forms a reflective subcontext of the context of sound logics. There is a restriction passage from logics to sound logics. With restriction and inclusion, the context of sound logics forms a coreflective subcontext of the context of logics. A composite logic, the abstract representation of the (local) logics of the theory of Information Flow (Barwise and Seligman, 1997), consists of a base logic and a sound logic sharing the same language and theory, where any sentence satisfied by the base logic structure is also satisfied by the sound logic structure. Composite logics form a context with two projective component passages to general logics and sound logics. Information systems (see below) can be defined in either theories, logics, sound logics or composite logics. In the sketch institution **Sk**, the category-theoretic approach to ontological specification (discussed in detail below), a sound logic is a sketched interpretation (generalized universal algebra) consisting of a sketch and an interpretation that satisfies that sketch. In the logical environment **IFC**, the basic institution for Information Flow (defined below), a logic consists of a classification (structure) and a theory sharing a common set of types (language); in addition, there is a subset of instances, called normal instances, which satisfy all sequents (sentences) in the theory. The local logics of Information Flow are the composite logics of **IFC** with the classification and theory providing the base logic component and restriction to the normal instances providing the sound logic component.

Institutions, which represent logical systems, abstract and generalize the semantic definition of truth (Goguen and Burstall, 1992), which consists of a relation of satisfaction between structures and sentences. In short, the institutional approach is abstract model theory. The first step of abstraction in the institutional approach is to regard each structure as an instance, each sentence as a type and each occurrence of satisfaction as a classification: a structure satisfies a sentence when the sentence-as-type classifies the structure-as-instance. In this regard, the theory of institutions and the theory of Information Flow are very compatible; indeed, one can regard the theory of institutions as an indexed or parametric theory of information flow, with each institution (the parameter) having a theory of information flow constructed over it and links between institutions (parameter map) providing comparisons between these theories of information flow. The second step of abstraction in the institutional approach is to regard the logical language, which provides the context for an ontology, to be an indexing object. The institutional approach refers to such an indexing object as a language (signature in Goguen and Burstall, 1992). The language of an institution often contains nonlogical symbols for constants, functions and relations. However, in the institutional approach such a detailed description is inessential, and hence is ignored in the abstraction. In summary, each ontology is indexed by a language and described by a classification.

The representational power of the institutional approach comes from the linkages between objects, such as languages and classifications. Languages are linked by language morphisms, which typically map the constant, function and relation symbols of one language to the constant, function and relation symbols of another language. A classification has instance and type collections and a classification relation between the two. For example, cars are classified by the combination structure-make-year — a particular car is an instance, and the combination 'Honda-Civic-1987' is a type. Classifications are linked by infomorphisms, which map between instance collections in the reverse direction (the instance map is said to be contravariant), map between type collections in the forward direction (the type map is said to be covariant), and require invariance of classification: a type classifies at the origin the image of an instance if and only if the image of the type classifies at the destination the instance. Just as languages index classifications in the institutional approach, so also language morphisms index infomorphisms. When two ontologies are indexed by languages linked by a language morphism, and described by two classifications, then the language morphism indexes an infomorphism that links the two classifications, thereby relating the two ontologies by invariance of satisfaction (invariance of truth under change of notation).

## 1.2. Semantic Integration

An information system (Barwise and Seligman, 1997) is a diagram of ontologies, where each ontology is represented as a logic or a theory. Since each logic (theory) has an underlying structure (language), an information system has an underlying distributed system, which is a diagram of structures (or languages). A channel over a distributed system is a corelation that covers the system (satisfies its semantic alignment constraints) with its vertex called the core of the channel. The optimal channel over a distributed system with sum vertex is an optimal (most refined) covering corelation. Semantic integration** (Kent, 2004) (Goguen, 2006) is the two-step process of alignment and then closure. The *alignment* of a distributed collection of ontologies is a human-oriented and creative process that builds a suitable information system. Alignment is called intersection in the chapter by Johnson and Rosebrugh. The *closure* of the information system is an automatic process of information flow. Closure has three phases: the first two phases are called unification (i) direct flow of the distributed system along the optimal channel (summing corelation over the underlying distributed system), and (ii) meet expansion of the direct flow image within the lattice of logics (theories) indexed by the sum distributed system; the last phase is called projective distribution (iii) inverse flow back along the same optimal channel. Unification is called theory blending in the chapter by Healy. For further details on the concepts discussed here, see the chapter by Kalfoglou and Schorlemmer.

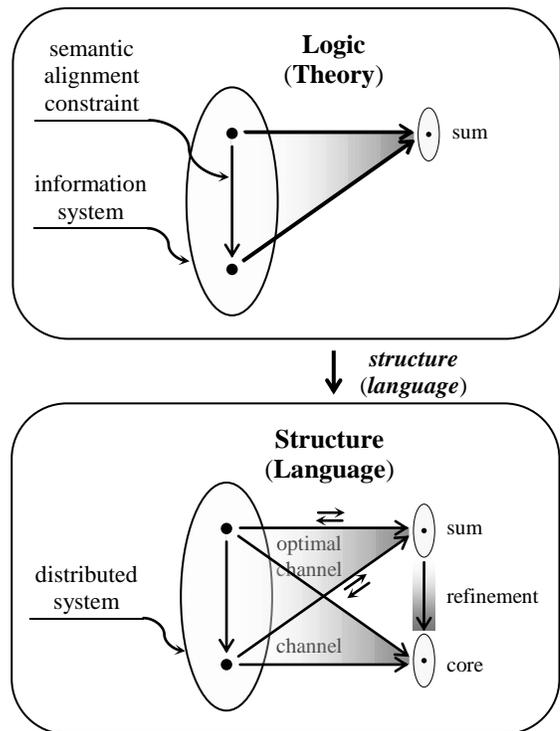

**Figure 1: Semantic Integration**

The logics (or theories) in the alignment diagram represent the individual ontologies, and the links between logics (or theories) in the alignment diagram represent the semantic alignment constraints. The alignment diagram is an information system with individual logics (or theories) representing parts of the system. The closure of an information system may be relative (partial) or absolute (complete) — relative closure is defined along an indexing passage, whereas absolute closure is defined along the unique indexing passage to the terminal indexing context. Absolute closure can be approximated by indexing passage composition. The relative sum information system along an indexing passage represents the whole system in a partially centralized fashion with the target indexing context defining the degree of centralization. The sum logic in the construction of the absolute closure of an information system represents the whole system in a centralized fashion, whereas the original information system and its closure represent the whole system in a distributed fashion. Ignoring the semantic constraints in the closure information system, the absolute closure is called the distributed logic of the information system in (Barwise and Seligman, 1997) (see also the chapter by Kalfoglou and Schorlemmer); but it is probably best to recognize it as an information system in its own right and to understand the properties of the closure operator.

Since logics (theories) over the same language are ordered by entailment, information systems with the same indexing context are ordered pointwise by entailment. Two information systems are pointwise entailment ordered when the component logics at each index are entailment ordered. Two information systems are ordered by system entailment when the closure of the first information system is pointwise entailment below the second information system. Hence, the following analogies hold between theories and information systems: theory closure ↔ system closure; reverse subset order for theories ↔ pointwise entailment order for systems; and theory entailment ↔ system entailment. A very important problem in distributed logic is the understanding of how one part of a system affects another part. System closure provides a solution. System closure, which is a closure operator with respect to reverse pointwise entailment order, describes how a distal part (ontology) of the system constrains a proximal part (ontology) of the system.

Figure 1 provides a graphic representation of semantic integration: in the Logic (Theory) context information systems are represented as ovals, ontologies are represented as nodes within ovals, and alignment constraints between ontologies are represented as edges within ovals; and in the Structure (Language) context distributed systems are represented as ovals, structures (languages) are represented as nodes within ovals, and channels are represented as triangles. The detailed theory of semantic integration in logical environments via system closure is developed in (Kent, 2008). A simple example of semantic integration, (analogous to one discussed in the chapter by Kalfoglou and Schorlemmer), starts with a collection of two ontologies represented as theories: (1) the alignment step might create a third bridging, mediating or alignment theory, whose types represent equivalent pairs of types in the two original ontologies, with two mappings from these mediating types back to the types in the equivalent pairs — the alignment diagram (information system) would then have indexing context generated by an inverted vee-shaped graph with three nodes and two edges; (2) the two phases in unification create a sum theory, and the projective distribution phase creates the absolute system closure information system (this corresponds to the integration theory mentioned in the chapter by Kalfoglou and Schorlemmer). For a more general and heterogeneous (across logical systems) approach to semantic integration, see the paper (Schorlemmer and Kalfoglou, 2008).

## 1.3. Architecture

Here we give a quick overview of the architecture (unifying or coherent form or structure) of the institutional approach; later sections provide more detail. Most of the sections in this chapter discuss elements illustrated in Figure 2. Where these and related elements are defined in this chapter, they are italicized. The architecture of the institutional approach to the theory and application of ontologies is illustrated in this figure. The first thing that we see is the separation of the architecture into a general and a special theory. Dual notions (reversed linkage orientation) exist in both general and special theories. The three basic ideas in the general theory are contexts (rectangle), indexed contexts (large triangle) and their indexed fibers (lens-shaped figure) and diagrams (small triangle). The two basic processes (passages) in the general theory are coalescence and fusion. Coalescence is composition. For an ambient context **V**, coalescence operates on **V**-diagrams and **V**-indexed contexts and returns index contexts (these terms are defined below). Here **V** is a valuation context for diagrams and an indexing context for indexed contexts; coalescence bonds diagrams to indexed contexts. In the special case where **V** is the context classification **Cls** (**Cls**-diagrams are institutions), by holding the institution (**Cls**-diagram) fixed and letting S denote its indexing language context, we can regard coalescence as a map from the core context of **Cls**-indexed contexts to the core context of S-indexed contexts; for example, mapping the logic indexed context for classifications to the logic indexed context for the fixed institution (by applying fusion to this process, we define a logic map from the core diagram in Figure 6 to the core diagram in Figure 6a). Fusion or the Grothendieck construction (Grothendieck, 1963) is a way for homogeneously handling situations of structural heterogeneity (Goguen, 2006). It maps the heterogeneous situations represented by indexed contexts to the homogeneous situations represented by contexts. There are two kinds of fusion. The basic fusion process operates on indexed contexts and returns contexts. The derived fusion process has two stages: coalescence, then basic fusion. There are two versions of the general theory architecture (Figure 2), one the normal version and the other the dual version. These are linked by the three involutions (see the discussion on duality below) for contexts, indexed contexts and diagrams. Because these involutions are isomorphisms, the normal and dual processes for coalescence, basic fusion and derived fusion can be defined in terms of one another. However, the dual versions have simpler definitions from more basic concepts. All notions in the architecture for the institutional approach are 2-dimensional notions, having not just links between objects but also connections between links (Figure 4).

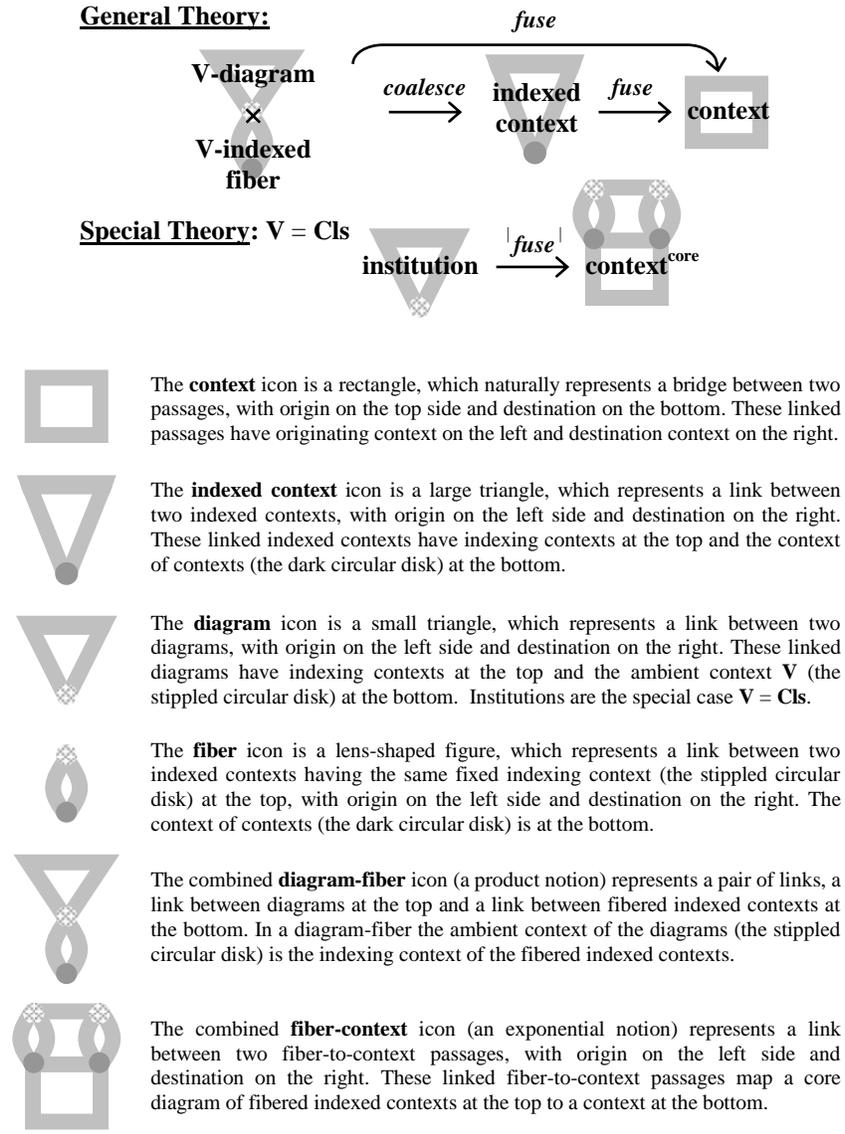

**Figure 2: Architecture**

The special theory fixes the ambient context to be the context of classifications $\mathbf{V} = \mathbf{Cls}$ (This equality represents assignment of the constant context $\mathbf{Cls}$ to the variable context $\mathbf{V}$). The context $\mathbf{Cls}$ is the central context in the theories of Information Flow and Formal Concept Analysis (Ganter and

Wille, 1999). The two basic ideas in the special theory are context$^{core}$s (rectangle with lens-shaped top edge) and institutions (small triangle). The context of context$^{core}$s is a subcontext of (core-shaped diagram within) the context of passages from **Cls**-indexed contexts to contexts (the **Theory** node in the core diagram (Figure 6) is the fusion of one example of a **Cls**-indexed context that maps a classification to its context of theories and maps an infomorphism to its theory passage via inverse flow). Hence, the context of context$^{core}$s is an exponential context — a product-exponent adjoint currying operator (upper corners) is used on fusion in the general theory before specializing. Connections (links of links; Figure 4) in context$^{core}$ are called modifications. Institutions are **Cls**-diagrams, diagrams in the ambient context of classifications. The basic process in the special theory is fusion: the fusion of an institution is a passage from the context of context$^{core}$s to the context of contexts. The core-shaped fusion diagram for an institution (Figure 6a) is embeddable into the universal core-shaped fusion diagram based only on the context of classifications (Figure 6). However, this universal core-shaped fusion diagram (Figure 6) is actually just one of the core-shaped fusion diagrams (Figure 6a) gotten by using the terminal institution consisting of the identity classification passage $1_{\mathbf{Cls}}$. Links between embeddings is coherently connected through modifications. Although potentially large, the core diagram shape is actually quite small for practical purposes. And it is different depending on duality. Just like the general theory, there are two versions of the special theory architecture, normal and dual, and these are also linked by involutions, a diagram involution for institutions and a composite (core and context) involution for context$^{core}$s. The normal version

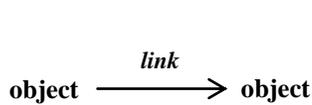
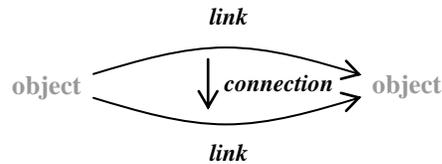

**Figure 3: Link**  **Figure 4: Connection**

(institutions and institution morphisms) has a naturally defined core diagram (Figure 6a) consisting of structures, theories and logics, plus linking and connecting elements. The dual version (institutions and institution comorphisms) has a simple core diagram consisting only of theories. The advantage of the normal version is the existence of (local) logic contexts (the **Logic** node in the core-shaped fusion diagram for an institution (Figure 6a) is the fusion of one example of a indexed context that maps a language to its logic order and maps a language morphism to its logic order map). The advantage of the dual version is the cocontinuity of theory passages (they preserve sums of theories). For either notion of fusion (normal or dual), the theory index in the core represents the lattice of theories construction as a passage from the context of institutions (normal version) or the context of coinstitutions (dual version) to the context of contexts (the context of coinstitutions has institutions as objects and institution comorphisms as links).

2. CONTEXTS

*2.1. General Theory*

Both languages and classifications are objects of a mathematical context. A *mathematical context* (Figure 2) corresponds to some species of mathematical structure (Goguen, 1991) (see the Chapter by Healy for a more detailed discussion). It consists of a collection of objects, a collection of links relating one object to another, a way to compose links into new links, and a special identity link for each object. A link in a context (Figure 3) has an orientation or direction; that is, it begins or originates at one object and ends or has destination at another. The link relates the beginning or originating object with the ending or destination object. Two links are composable when the destination of one is the origin of the other. A link in a context is an isomorphism when there is another bicomposable link, where the two compositions are identity. Any order is a context with the orderings being the links. A 2-dimensional mathematical context also has a collection of connections (Figure 4), which relate one link to another link – connections are links between links. There are vertical and horizontal ways to compose connections into new connections. The horizontal way corresponds to link composition (see the discussion of 2-categories in Healy's chapter).

A *passage* (Figure 3a) from a beginning or originating context to an ending or destination context consists of a map from the beginning object collection to the ending object collection and a map from the beginning link collection to the ending link collection that preserves linkage direction and composition. Passages themselves can be composed. Two passages are composable when the destination of one is the origin of the other. The composition passage is defined coordinate-wise: compose the object maps and compose the link maps. An order map is a passage between two orders with order-preservation representing the link map. A 2-dimensional passage between 2-dimensional contexts also has a map of connections that preserves connecting direction and vertical and horizontal composition. For any map beginning in one collection and ending in another collection, a fiber over a fixed item in the destination collection is the collection of all elements in the beginning collection that map to that fixed item. Similarly, for any passage, a fiber over a fixed object in the ending context is a subcontext of the beginning context consisting of the collection of all objects that map to that fixed object and the collection of all links that map to the identity link on that fixed object. For a 2-dimensional passage, the fiber contains the collection of connections that map to the identity connection on that identity link.

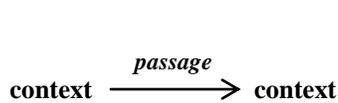

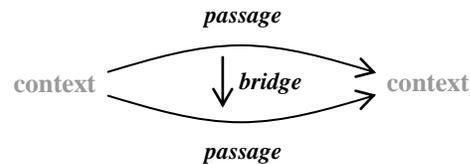

**Figure 3a: Passage**     **Figure 4a: Bridge**

A *bridge* (Figure 4a) from a beginning or originating passage to an ending or destination passage (between the same mathematical contexts) consists of a map from the originating context's object collection to the destination context's link collection, which naturally preserves linkage. These links in the destination context are called components of the bridge. Passages and bridges can be (horizontally) composed when the destination context of one is the originating context of the other. In the passage-bridge composition, the object that indexes a component is initially mapped by the passage. In the bridge-passage composition, the component (that is indexed by an object) is finally map by the passage. Bridges themselves can be composed, both vertically and horizontally. Two bridges are vertically composable when the destination of one is the origin of the other. The vertical composition bridge is defined component-wise: for each object in the originating context, compose the components in the destination context. Vertical composition is orthogonal to passage composition. Two bridges are horizontally composable when the destination context of one is the originating context of the other. The horizontal composition bridge has two equal and interchangeable definitions. One is the vertical composition of the composition of the first bridge and the originating passage of the second bridge and the composition of the destination passage of the first bridge and the second bridge. The other has a dual definition. Horizontal composition is parallel to passage composition. Any ordering on a pair of order maps is a bridge. Bridges enrich the context of contexts into a 2-dimensional mathematical context with contexts as objects, passages as links and bridges as connections.

Philosophically, the notion(s) of identity takes on several forms. Two objects in a context are equal up to identity when they are the same; they are equal up to isomorphism when they are linked by an isomorphism; and they are equal up to morphism when they are linked. Two contexts are identical when they are equal; they are isomorphic when they are linked by an isomorphism: and they are equivalent when they are linked by an equivalence.

An *invertible passage* (*equivalence*) or inverse pair of passages from a beginning or originating context to an ending or destination context is a pair of bicomposable passages, a left passage in the same direction and a right passage in the opposite direction, which are generalized (relaxed) inverses for each other in the sense that the compositions are identity naturally up to (iso)morphism. This means that the identity passage at the beginning context is connected to the left-right composite by a bridge called the unit, and that the right-left composite is connected to the identity passage at the ending context by a dual bridge called the counit. Unit and counit can be composed with the left and right passages. Unit and counit bridges are coherently related by two "triangle equalities" that are dual to each other: the vertical composition of the unit-left (right-unit) composite with the left-counit (counit-right) composite is the identity bridge at the left (right) passage. The context of invertible passages has contexts as objects and invertible passages as links. There are left and right projections from the context of invertible passages to the context of contexts with left being covariant and right being contravariant.

Duality (Figure 5) is important in the institutional approach to ontologies, and can be confusing if not approached with some caution. In the institutional approach, there are several kinds of duality at work. The opposite of a context flips the direction of its links. The opposite of a passage has the same action, but maps a flipped link to the flip of the image link. The opposite of a bridge has the same components, hence has its direction flipped. The opposite of an invertible passage applies the opposite to all of its components, left, right, unit and counit.

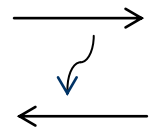

**Figure 5: Involution**

It flips its direction and changes dual notions: the left (right, unit, counit) of the opposite is the opposite of the right (left, counit, unit). A contravariant passage is a (covariant) passage from the opposite of a context — same action, but change of perspective.

Duality leads to some important involutions in the institutional approach, which are defined on the basic ideas in Figure 2 — contexts, indexed contexts and diagrams. The context involution maps a context, passage or bridge to its opposite. It is a 2-dimensional isomorphism on the context of contexts, which is covariant on passages and contravariant on bridges. The indexed context involution maps a dual indexed context, a dual indexed link or a dual indexed connection to itself, but changes from a contravariant to a covariant perspective. It is a 2-dimensional isomorphism between the context of dual indexed contexts and the context of indexed contexts, which is covariant on dual links and contravariant on dual connections. There is a fibered version of this: for any ambient context **V**, there is a **V**-indexed context involution from the context of dual **V**$^{op}$-indexed contexts to the context of **V**-indexed contexts, where **V**$^{op}$ is the opposite of **V**. For any ambient context **V**, the diagram involution maps a **V**$^{op}$-diagram or **V**$^{op}$-diagram colink to its opposite **V**-diagram or **V**-diagram link, defined by flipping its components, either indexing context and passage or indexing passage and bridge. It is an isomorphism between the context of **V**$^{op}$-codiagrams and the context of **V**-diagrams, which is covariant on morphisms.

Several constructions beyond duality are also used in the institutional approach. The product of two collections contains all pairs of elements, where the first (second) element is from the first (second) collection. Similarly, the *product* of two contexts contains all pairs of objects and all pairs of links from the component contexts, which preserve products of origins and destinations. The two original contexts are called the components of the product. The exponent of two collections has all the maps between the collections as elements. Similarly, the *exponent* of two contexts has all the passages between the contexts as objects, the bridges between these passages as links, and vertical composition of these bridges as composition. Finally, any passage with a product origin has an *adjoint* form, which is a passage from one product component to the exponent of the other product component and the destination of the passage.

*2.2.    Special Theory*

The context of classifications **Cls** has classifications as objects and infomorphisms as links[†]. A classification has instance and type collections and a classification relation between the two. Classifications are linked by infomorphisms, which map between instance collections in the reverse direction (the instance map is said to be contravariant), map between type collections in the forward direction (the type map is said to be covariant), and require invariance of classification: a type classifies at the origin the image of an instance if and only if the image of the type classifies at the destination the instance. There are two component projection passages, instance and type, from the context of classifications to the context of sets; instance is contravariant and type is covariant.

A theory of a classification is a subset of types. For any classification, the intent of an instance is the (theory) subcollection of types that classify the instance, and dually the extent of a type is the subcollection of instances classified by that type. A theory classifies an instance of a classification when each type in the theory classifies the instance. Thus, any classification lifts to its theories; that is, any classification has an associated instance-theory classification, where types are replaced by theories. The extent of a theory is the subcollection of all instances classified by that theory; that is, the intersection of the extents of the types in the theory. A theory entails a type when any instance classified by the theory is also classified by the type; that is, when the extent of the theory is contained in the extent of the type. Theories are linked by type maps that preserve entailment. The collection of all types entailed by a theory is called the closure of the theory. Closure is an operator on theories. Types in the theory are called axioms, whereas types in the closure are called theorems. A (local) logic of a classification has two components, a theory and an instance, with a common underlying language.

For any classification, the three collections of instances, theories and logics are ordered. Two instances are ordered when any type that classifies the second instance also classifies the first instance; that is, when the intent of the first contains the intent of the second. Two theories are ordered when any axiom of the second is a theorem of the first; that is, when any type in the second theory is entailed by the first theory; that is, when the closure of the first theory contains the (closure) of the second theory; that is, when the extent of the first is contained in the extent of the second. Two logics are ordered when their instance and theory components are ordered. In any classification, the intent of an instance is maximal in subset order and minimal in entailment order. This defines a maximal theory map from instances to theories.

Give any map, direct image and inverse image form an invertible pair of maps between subsets that preserves inclusion. For any infomorphism, the (covariant) direct flow on theories is the direct image of the type function and the (contravariant) inverse flow on theories is the composition of closure followed by the inverse image of the type function. For any infomorphism, the type function preserves entailment: if any theory entails a source type at the origin, then the direct flow of the theory entails the image of the type at the destination. When the instance function of the infomorphism is surjective, the type function allows borrowing (useful in theorem-proving): any theory entails a type at the origin if and only if the direct flow of the theory entails the image of the type at the destination. Any infomorphism lifts to its theories, where the type map is replaced by the direct flow map. Lifting to theories is a passage on the context of classifications. For any infomorphism, the (contravariant) logic map, which maps between logic collections in the reverse direction, has two components, the inverse flow and the instance map. For any infomorphism, the instance map preserves order on instances, the direct and inverse flow preserve (entailment) order on theories, and the logic map preserves order on logics. Direct and inverse flows form an invertible map on theories. They also form an invertible map on logics. Direct flow on logics preserves soundness; the direct image of a sound logic is also sound. Inverse flow on logics preserves completeness; the inverse image of a complete logic is also complete. Direct and inverse flow on logics forms the basis of the theory of Information Flow.

The context of classifications is fundamental in the institutional approach, and it is (co)complete. But the context of concept lattices is equivalent to it (Kent, 2002), and the context of concept orders is pseudo-equivalent to it. Hence, these also are cocomplete. A concept order (complete order with two-sided generators) consists of an order with all meets and joins, an instance set, a type set, a map that embeds instances (types) as order elements, such that any element is the join (meet) of some subcollection of embedded instances (types). When a concept order satisfies antisymmetry (isomorphic elements are identical) it is called a concept lattice. A concept morphism links concept orders. It consists of an invertible pair of order maps called the left (right) inverse, and a map linking the instance (type) collections in the opposite (same) direction, where the left (right) inverse preserves embedded instances (types). Compositions and identities are defined component-wise. The context of concept orders (lattices) has concept orders (lattices) as objects and concept morphisms as links. The context of concept lattices is a full subcontext of the context of concept orders. Each of the three contexts (classifications, concept lattices and concept orders) comes equipped with, and is definable in terms of, component projection functors. The 2-dimensional diagram consisting of these three contexts, along with their connecting passages and bridges, is called the conceptual core.

3. INDEXED CONTEXTS

*3.1.    General Theory*

An index is a pointer used to indicate a value. A map or list is the simplest mathematical representation for an index, mapping an indexing set to a set of items of a certain type. A passage is a structured representation for an index, mapping an indexing context to a context of objects of a certain type. A structured index of a certain type *X* is describe as an indexed *X*. An *indexed context* (Figure 2) is a passage into the context of contexts. As such, it is a special kind of diagram. It originates at an indexing context, maps indexing objects to component contexts, maps indexing links to component passages, inverts direction (it is contravariant) and preserves composition and identities up to isomorphism. An indexed order is the special case of a passage into the context of orders – the passage maps indexing objects to component orders, and maps indexing links to component order maps. The involute of an indexed context is the composition of the indexed context with the context involution. A *dual indexed context* is the same; except it preserves direction (it is covariant). The indexed involution maps a dual indexed context to an indexed context by flipping the indexing context to its opposite and changing the perspective from covariance to contravariance.

An *inversely-indexed context* (corresponding to the notion of a locally reversible indexed context in Tarlecki, Burstall and Goguen, 1991) is a passage into the context of invertible-passages. It also originates at an indexing context, mapping indexing objects to component contexts and indexing links to component invertible passages, inverting direction, and preserving composition and identities up to isomorphism. There are two components, a left component indexed context and a right component dual indexed context. We think of the modifiers "inversely" and "locally reversible" as applying to the left component indexed context; thus, inversely indexed contexts are special indexed contexts with right inverses. We define cocompleteness for dual indexed and inversely-indexed contexts. A dual index context is component-complete when all component contexts are complete. It is component-continuous when it is component-complete and all the component passages are continuous. It is cocomplete when it is component-continuous and the indexing context is cocomplete. An inversely indexed context is component-complete (component-continuous, cocomplete) when its right inverse dual indexed context is so.

An *indexed link* from a beginning or originating indexed context to an ending or destination indexed context consists of an indexing passage from the indexing context of origin to the indexing context of destination and a bridge from the beginning passage to the composition of the opposite of the indexing passage with the ending passage. Indexed links can be composed. Two indexed links are composable when the destination of one is the origin of the other. The composition of a composable pair is defined by the composition of their indexing passages and the vertical composition of their bridges. An indexed order map between indexed orders is the special case where the bridge components are order maps. A *dual indexed link* is the same, except that it links dual indexed contexts. The indexed involution maps a dual indexed link to an indexed link by flipping the indexing passage to its opposite and changing the perspective from covariance to contravariance. We define cocontinuity for only dual indexed links. A dual indexed link is component-continuous when it links component-continuous dual indexed contexts, and the component passages of its bridge are continuous. It is cocontinuous when it links cocomplete dual indexed contexts, it is component-continuous, and its indexing passage is cocontinuous.

An *indexed connection* from a beginning or originating indexed link to an ending or destination indexed link consists of an indexing bridge from the beginning indexing passage to the ending indexing passage, which preserves bridging up to morphism in the sense that the beginning bridge is linked to the vertical composition of the ending bridge with the opposite of the indexing bridge. Indexed connections can be composed by vertical composition of their indexing bridges. A *dual indexed connection* is the same, except that it links dual indexed links. The indexed involution contravariantly maps a dual indexed connection to an indexed connection by flipping the indexing bridge to its opposite and changing the perspective from covariance to contravariance.

The context of indexed contexts is a 2-dimensional context, whose objects are indexed contexts, whose links are indexed links, and whose connections are indexed connections. The context of indexed-orders is a special case. Similar comments hold for the dual notions. The indexed involution is a 2-dimensional isomorphism from the context of dual indexed contexts to the context of indexed contexts, which is covariant on indexed links and contravariant on indexed connections. There is a 2-dimensional indexing passage from the context of indexed contexts to the context of contexts, which maps an indexed context to its indexing context, maps an indexed link to its indexing passage, and maps an indexed connection to its indexing bridge. Also, there is a 2-dimensional indexing passage from the context of indexed orders to the context of contexts.

In the institutional approach we are interested in the fibers for the 2-dimensional indexing passage from the context of indexed orders to context of contexts. The *indexed fiber* (Figure 2) over a fixed context consists of the following: the objects are the indexed orders with that fixed context as indexing context, the links are the indexed order maps with the identity passage on that fixed context as indexing passage, and the connections are the indexed connections with the identity bridge on that identity passage as indexing bridge. Hence, an indexed link in a fiber, called an indexed order map, consists of a bridge between origin and destination indexed orders, and an indexed connection in a fiber, called an order pair of indexed order maps, consists of an ordering between origin and destination indexed order maps. We have inversely-indexed orders in fibers. Furthermore, we can define dual versions of all these notions. Finally, we define an additional notion in a fiber: an indexed invertible pair of order maps is a pair of bicomposable indexed order maps, a left indexed order map in the same direction and a right indexed order map in the opposite direction, which are generalized (relaxed) inverses for each other in the sense that the bridge components are invertible pairs of order maps; that is, vertical composition of the bridges is unique up to order both ways.

## 3.2. Special Theory

Here we discuss the core-shaped universal diagram in the special theory of the architecture. Consider the fiber of indexed orders over the fixed context **Cls**. Instance is an indexed order that maps a classification to its order of instances and maps an infomorphism to its instance order map. Inverse flow is an indexed order that maps a classification to its theory (entailment) order and maps an infomorphism to the inverse flow of its type map. Logic is an indexed order that maps a classification to its logic order and maps an infomorphism to its logic order map. Direct flow is a dual indexed order that maps a classification to its theory (entailment) order and maps an infomorphism to the direct flow of its type map. This dual indexed order is component-complete, since for any classification the component theory order is a complete lattice. It is component-continuous, since for any infomorphism the component direct image passages are continuous, being right inverses of inverse image. The direct flow dual indexed order is cocomplete, since it is component-continuous and the indexing context of classifications is cocomplete. Inverse and direct flow form an inversely-indexed order with inverse flow as left component and direct flow as right component. This inversely-indexed order is component-complete (component-continuous, cocomplete), since its direct image right adjoint dual indexed context is so. There is a maximal theory indexed order map between the instance and inverse flow indexed orders, whose bridge has the maximal theory maps as its components. From the logic indexed order, there are two projection indexed order maps to the instance and the inverse flow indexed orders. The normal core diagram shape has three nodes or index orders (instance, inverse flow and logic) and three edges or indexed order maps (maximal theory and logic projections). Its fusion (Figure 6), a core diagram in the context of contexts, has four contexts (classification, instance, theory and logic) and five passages (projections, base passages, and maximal theory). The instance projection passage within the core diagram for classifications, just like the structure projection passage within the core diagram for an institution (Figure 6a), is the lifting of its base passage originating from the context of theories. The dual core diagram shape has one node or dual index order (direct flow) and no edges. Its fusion is just the context of theories and the context of classifications with a base passage in between. A link in the context of instances is an infomorphism that maps the instance at the destination to a specialization of the instance at the origin. An instance is more specialized that another when it is classified by more types. A link in the context of theories is an infomorphism that maps the theory at the destination to a specialization of the theory at the origin; equivalently, maps the theory at the origin to a generalization of the theory at the destination. A theory is more specialized that another when any axiom of the second is a theorem of the first. A link in the context of logics is an infomorphism that is both an instance link and a theory link.

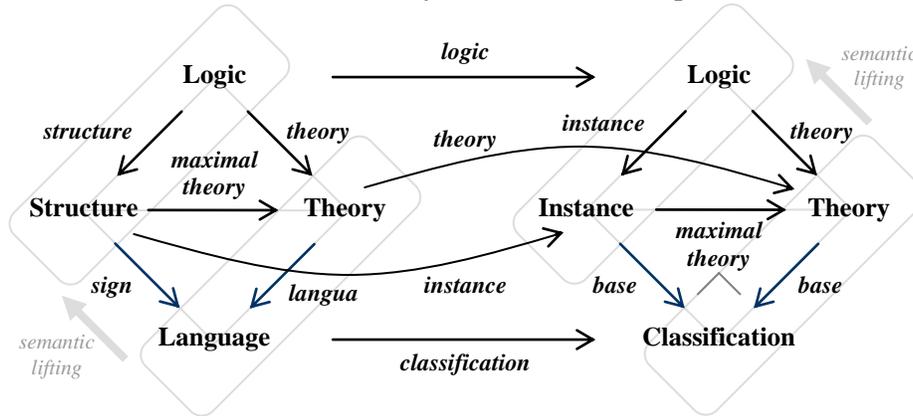

Figure 6a: Core Diagram (institutions)   Figure 6: Core Diagram (classifications)

## 4. DIAGRAMS

### 4.1. General Theory

Let **V** be any context within which we will work. Of course, one would normally choose a context **V** that has some useful properties. We keep that context fixed throughout the discussion and call it the ambient context. We regard the objects and links in the ambient context **V** to be values that we want to index, and we focus on a particular part of the ambient context **V**. We use a passage into **V** for this purpose. A *diagram* (Figure 7) is a passage from an indexing context into the ambient context **V**. The objects in the indexing context are called indexing objects and the links are called indexing links. A diagram may be presented by using a directed graph to generate an indexing path context. If so, such a graph is usually called the shape of the diagram. By extension, the indexing context of any diagram can be called its shape. The diagram involution maps a $\mathbf{V}^{op}$-diagram to its opposite **V**-diagram, defined by flipping its indexing context and passage.

Diagrams can be linked. A *diagram link* (Figure 8) from a beginning or originating diagram to an ending or destination diagram consists of a passage between indexing contexts called an indexing passage and a bridge from the composition of the indexing passage with the destination passage to the beginning passage. Two diagram links are composable when the destination of one is the origin of the other. The composition diagram link is defined coordinate-wise: compose the indexing passages and vertically compose the bridges. Diagrams can also be linked in a dual fashion. A *diagram colink* from a beginning or originating diagram to an ending or destination diagram consists of an indexing passage as before, but a bridge in the opposite direction: from the beginning passage to the composition of the indexing passage with the destination passage. Composition is defined similarly. The diagram involution maps a $\mathbf{V}^{op}$-diagram colink to its opposite **V**-diagram link, defined by flipping its indexing passage and bridge. There is a context of **V**-diagrams with diagrams as objects and diagram links as links, and there is a context of **V**-codiagrams with diagrams as objects and diagram colinks as links. There is an indexing passage to the context of contexts from the context of **V**-(co)diagrams, which maps a diagram to its indexing context and maps a diagram (co)link to its indexing passage. The diagram involution is an isomorphism between the context of $\mathbf{V}^{op}$-codiagrams and the context of **V**-diagrams, which is covariant on morphisms. There is a terminal diagram consisting of the identity passage on the ambient context **V**, so that **V** is the indexing context. From any diagram there is a trivial diagram (co)link to the terminal diagram with the indexing passage being the passage of the diagram and the bridge being identity.

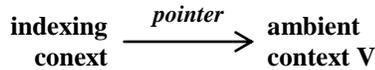
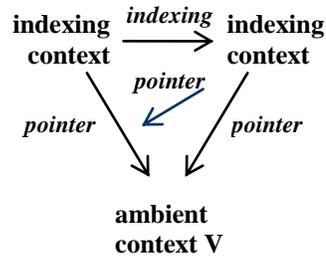

**Figure 7: Diagram**  **Figure 8: Diagram Link**

Although dual, the links and colinks between diagrams seem to be independent. However, a strong dependency exists when their indexing passages are invertible. Such a strongly dependent pair is called a duality. More precisely, a *duality* is a pair consisting of a diagram colink and a diagram link whose indexing passages form an invertible pair of passages with the left component the indexing passage for the diagram colink and the right component the indexing passage for the diagram link. This implies that colink and link are between diagrams in opposite directions. Then colink and link are definable in terms of each other: (loosely) the bridge of the colink is the vertical composition of the unit with the bridge of the link and the bridge of the link is the vertical composition of the bridge of the colink with the counit.

In the institutional approach we are interested in the fibers for the indexing passage from the context of **V**-codiagrams to context of contexts. The indexed fiber over a fixed context consists of the following: the objects are the **V**-diagrams with that fixed context as indexing context, and the links are the **V**-diagram colinks with the identity passage on that fixed context as indexing passage. If we think of the indexing passage as a means of moving diagrams along indexing contexts, then a fiber link is one with an identity indexing passage. That is, a fiber link is just a bridge between two passages with the same shape that map into the ambient context **V**; it is a bridge from the passage of origin to the passage of destination. A constant diagram is a diagram that maps all indexing objects to a particular object in **V** and maps all indexing links to the identity on that object of **V**. For any object in **V** and any context (as shape), there is a constant diagram over that object with that shape. A *corelation* is a fiber link to a constant destination diagram. A constant diagram link is a fiber link between two constant diagrams (with the same shape) whose bridge components are all the same – a particular link in **V** between the objects of the constant diagrams. For any link in **V** and any context (as shape), there is a constant diagram link over that link with that shape.

For any diagram, a summing corelation is an initial corelation originating from that diagram: any other corelation originating from that diagram is the vertical composition of the summing corelation with the constant diagram link over a unique link in **V**. Any two summing corelations for the same diagram are isomorphic, and hence conceptually identical. For a summing corelation the object is called a *sum* of the diagram and the component links are called sum injections. The sum of a diagram is a kind of constrained sum: disjointly sum the component objects indexed by the diagram, and constrain these objects with the links indexed by the diagram. The ambient context **V** is said to be *cocomplete* when sums exist for all diagrams into it. Relations, producting relations, products and completeness are defined dually. Two equivalent contexts have the same sums up to isomorphism – a sum in one context is isomorphic to a sum in an equivalent context. Two pseudo-equivalent contexts have the same sums up to equivalence – a sum in one context is equivalent to a sum in a pseudo-equivalent context. The context of **V**-diagrams is complete (cocomplete) when **V** is complete (cocomplete), and the context of **V**-codiagrams is complete (cocomplete) when **V** is cocomplete (complete) (Goguen and Roşu, 2002). Hence, the context of $\mathbf{V}^{op}$-codiagrams is complete (cocomplete) when $\mathbf{V}^{op}$ is complete (cocomplete). This is compatible with the fact that the context of $\mathbf{V}^{op}$-codiagrams is isomorphic to the context of **V**-diagrams via the diagram involution. There is an issue about the "smallness" of diagrams that we are ignoring here. Composition by a passage maps a diagram in the originating context (an originating diagram) to a diagram in the destination context (a destination diagram). A passage is *cocontinuous* when the passage maps the sum of any originating diagram to a sum of the corresponding destination diagram. Continuity of a passage has a dual definition using limits.

### *4.2. Special Theory*

An *institution* or logical system (Figure 7a) is a diagram in the ambient context of classifications **Cls**. The context of institutions is the context of **Cls**-diagrams, where links are called institution morphisms (Figure 8a). The context of coinstitutions is the context of **Cls**-codiagrams, where links are called institution comorphisms. There is a terminal institution, which is the terminal **Cls**-diagram; it consists of the identity classification passage $1_{\mathbf{Cls}}$. The ambient context **Cls** is both complete and cocomplete. Hence, the context of institutions and the context of coinstitutions are both complete and cocomplete, with the identity passage on **Cls** being the terminal institution. (Much of the theory of

Information Flow is based upon the fact that **Cls** is cocomplete — the distributed systems of Information Flow are represented as diagrams in **Cls**, channels covering such distributed systems are represented as corelations in **Cls**, and minimal covering channels are represented as sums in **Cls**.)

In more detail, an institution (Goguen and Burstall, 1992) consists of an indexing language context and a classification passage into the ambient context of classifications. Indexing objects for an institution are called languages or vocabularies, and indexing links are called language or vocabulary morphisms. The classification passage maps a language to a classification called the "truth classification" of that language (Barwise and Seligmann, 1997). For any language, the instances of its classification are called structures, the types are called sentences, and the classification relation is called satisfaction. For any structure and sentence in a satisfaction relationship, we say that "the structure satisfies the sentence" or "the sentence holds in the structure" or "the sentence is satisfied in the structure" or "the structure is a structure of the sentence". The classification passage maps a language morphism to an infomorphism, and the invariance of classification expresses the invariance of truth under change of notation: the image of a structure satisfies a sentence at the origin if and only if the structure satisfies the image of the sentence at the destination. A structure satisfies (is a model of) a theory when it satisfies all axioms of (sentences in) the theory. A theory entails a sentence when that sentence is true in all models of the theory. The closure of a theory is the collection of sentences entailed by the theory. Two theories are ordered by entailment when the first theory entails every axiom of the second theory; that is, when the closure of the first theory contains the second theory. Two models are entailment ordered when the theories that they induce are so ordered. Two logics are entailment ordered when their component models and theories are so ordered.

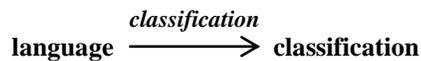
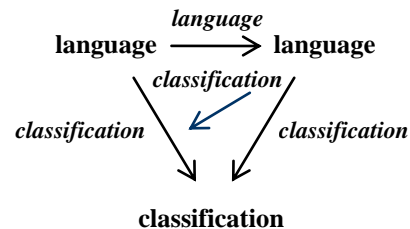

**Figure 7a: Institution**          **Figure 8a: Institution Morphism**

The equivalences between the context of classifications and the context of concept lattices or the context of concept preorders allows either of these to be equivalently used as the ambient context in defining the context of (co)institutions (Kent, 2004). That is, we can equivalently regard an institution to be a diagram of concept lattices. Here each language indexes a concept lattice, where the intent of a concept is a closed theory and concept order is reversed subset order on closed theories. Or, we can equivalently regard an institution to be a diagram of concept preorders. Here each language intentionally indexes a "lattice of theories" with each preorder element being a theory and order on theories being entailment order.

Examples of institutions include the following (Goguen and Burstall, 1992) (Goguen and Roşu, 2002) (Goguen, 2007): first order logic with first order structures as structures, many sorted equational logic with abstract algebras as structures, Horn clause logic, and variants of higher order and of modal logic. Other examples (Mossakowski, Goguen, Diaconescu and Tarlecki, 2005) of institutions include intuitionistic logic, various modal logics, linear logic, higher-order, polymorphic, temporal, process, behavioral, coalgebraic and object-oriented logics. Here are more detailed descriptions of some institutions.

In the institution **EQ** of equational logic (universal algebra), a language is a family of sets of function symbols, and a language morphism is a family of arity-preserving maps of function symbols. The set of sentences indexed by a language is the set of equations between terms of function symbols. The sentence translation function indexed by language morphism is defined by function symbol substitution. A structure of equational logic is an algebra, consisting of a set (universe) and a function for each function symbol in the language. Structure translation is a substitution passage; it is reduct with symbol translation. Satisfaction is as usual.

The institution **FOL** of unsorted first-order logic with equality extends the institution of equational logic by adding relation symbols. A language is a family of sets of function symbols as above, plus a family of sets of relation symbols with arity. A language morphism is a family of arity-preserving maps of function symbols as above, plus a family of arity-preserving maps of relation symbols. Sentences are the usual first order sentences. The set of sentences indexed by a language consists of closed first order formulae using function and relation symbols from the language. The sentence translation function indexed by a language morphism is defined by symbol substitution. Structures are the usual first order structures. A structure is a set (universe), a function for each function symbol (that is, an algebra, as above), and for each relation symbol a subset of tuples of that arity. Structure translation is reduct (substitution) with symbol translation. Satisfaction is as usual.

The theory of sketches (Barr and Wells, 1999) is a categorical approach for specifying ontologies. A sketch signature is a graph plus collections of cones and cocones each with an arity base in that graph. In an interpretation of a sketch the nodes of the underlying graph are intended to specify sorts (or types), the edges are intended to specify algebraic operations, and the cones (cocones) are intended to specify products (sums). A morphism between sketch signatures is a graph morphism between the underlying graphs of origin and destination, which preserves arity by mapping source cones (cocones) to target cones (cocones). Any context has an underlying sketch signature, whose graph is the underlying graph of the context (nodes are objects and edges are links), and whose cones (cocones) are the limiting cones (colimiting cocones) in the context. Given a sketch signature, a diagram in that signature is a diagram in the underlying graph; in a diagram any pair of paths with common beginning and ending nodes is called an equation. Given a sketch signature, an interpretation of that sketch signature in a fixed context (the context of finite sets is used in Johnson and Rosebrugh, 2007) is a sketch signature morphism into the underlying signature of the context, a graph morphism that maps cones (cocones) in the sketch to limiting cones (colimiting cocones) in the context. An interpretation morphism is a graph bridge from one interpretation to another. Given a sketch signature, an interpretation and a diagram in that signature, the interpretation satisfies the diagram when it maps the diagram to a commutative diagram in the fixed context; a commutative diagram in a context is a diagram where the composition of each side of an equation is equal.

A sketch consists of a sketch signature plus a collection of diagrams (equivalently, equations) in the underlying graph. The diagrams are intended to specify commutative diagrams in an interpretation. For any interpretation and sketch having the same signature, the interpretation satisfies the sketch, and is called a model (or an algebra) of the sketch, when it satisfies every diagram in the sketch. A homomorphism of models (algebras) is an interpretation morphism between models. Any interpretation defines a sketch (for which it is a model) having the same underlying signature and consisting of all diagrams satisfied by the interpretation. A sketch entails a diagram in its signature when any model of the sketch satisfies that diagram. Two sketches are ordered by entailment when they have the same underlying signature and any diagram of the second sketch is entailed by the first sketch. This is a preorder (reflexive and transitive). The closure of a sketch is another sketch with the same signature, but which contains all entailed diagrams. This is a closure operator on reverse entailment order (monotonic, increasing and idempotent). A sketch and its closure are isomorphic w.r.t.

entailment order. A sketch is closed when it is identical to its closure. A sketch morphism is a sketch signature morphism, which maps entailed diagrams of the first sketch to entailed diagrams of the second sketch.

In the institution `Sk` of sketches, a language is a sketch signature and a language morphism is a sketch signature morphism. The set of sentences indexed by a sketch signature is the set of diagrams in that signature. The sentence translation function indexed by a sketch signature morphism maps source diagrams to target diagrams by graph morphism composition. The set of structures indexed by a sketch signature is the set of interpretations of that signature. The structure translation function indexed by a sketch signature morphism is reduct (substitution). Satisfaction is defined above. Given any sketch signature morphism, the invariance of satisfaction expresses the invariance of truth under change of notation: the reduct of an interpretation satisfies a diagram at the origin if and only if the interpretation satisfies the translation of the diagram at the destination. There a several special kinds of sketch signatures useful for particular purposes: linear sketch signatures (or graphs) with neither cones nor cocones used, product sketch signatures (or multi-sorted equational logic languages) with only discrete cones and no cocones used, limit sketch signatures with any cones but no cocones used, and limit-coproduct sketch signatures (see the chapter by Johnson and Rosebrugh) with any cones but only discrete cocones used. Each of these special kinds of sketch signatures forms its own institution. In general, there is a trivial inclusion institution morphism from any one of these kinds to a more powerful kind; for example, from limit sketch signatures to general sketch signatures. Often in these special sketches, the (co)cones and graphs are required to be finite. The notion of a limit-product sketch is used to define the entity-attribute data model in (Johnson and Rosebrugh, 2007), which is an enriched extension of the traditional entity-relationship-attribute data model. The paper (Johnson and Rosebrugh, 2007) requires the model reduct passage to be a (op)fibration in order to define universal view updatability; a notion of cofibration is defined on sketch morphisms to ensure that this holds. It is an interesting question whether these notions have meaning and importance for an arbitrary institution.

Information systems for any institution of sketches can be defined over the context of sketches (theories), the context of models or algebras (sound logics), or even over a larger context of logics (not defined here). Within this institution a theory is a sketch and a theory morphism is a sketch morphism. This defines the context of theories. From this context there is an underlying passage to the context of sketch signatures. Any sketch signature has an associated context of interpretations and their morphisms. Any sketch signature morphism has an associated reduct (substitution) passage from target fiber context of interpretations to source fiber context of interpretations; this is defined by composition of the sketch signature morphism with interpretations and their morphisms. Hence, there is an indexed context from the context of sketch signatures (languages). The Grothendieck construction on this indexed context forms the context of structures (interpretations): an object is a pair consisting of a sketch signature and an interpretation having that signature; and a morphism is a pair consisting of a sketch signature morphism and a graph bridge from the originating interpretation to the reduct of the destination interpretation. From this context there is an underlying language passage (split fibration) to the context of sketch signatures. Any sketch has an associated a context of models (algebras) and their homomorphisms. For any sketch morphism, structure translation (reduct) preserves model satisfaction. Thus, any sketch morphism has an associated reduct passage between fiber contexts of models. Hence, there is an indexed context from the context of sketches (theories). See the model category functor in (Barr and Wells, 1999). Fusion (the Grothendieck construction) on this indexed context forms the context of sound logics (models or algebras): an object is a pair consisting of a sketch and a model (algebra) of that sketch; and a morphism is a pair consisting of a sketch morphism and a homomorphism from the originating model to the reduct of the destination model. From this context there are projection passages to the context of sketches and the context of interpretations.

A harmonious unification between the theories of institutions and Information Flow works best in a logical environment. A *logical environment* is a structured version of an institution, which takes the philosophy that semantics is primary. A logical environment requires a priori (1) the existence of a context of structures that is cocomplete and (2) the existence of a fibration (Cartesian passage) from the context of structures to the context of languages that factors through the order-theoretic and flatter context of structures built by fusion from just the underlying institution. An even more structured version of logical environment requires existence of a left adjoint to the fibration. The basic institution of Information Flow and an analogous institution of sorted first order logic are important examples of such logical environments.

The logical environment **IFC** is the basic institution for Information Flow. A structure is a classification, and a structure morphism is an infomorphism. A language is a set (of types) and a language morphism is a (type) function. The underlying language passage from the context of classifications to the context of sets is the type projection passage. A sentence is a sequent of types consisting of pairs of type subsets, antecedent and consequent. Sentence translation is direct image squared on types. A theory consists of sets of sequents; equivalently, a theory consists of a (type) set and an endorelation on subsets of types. A closed theory is known as a regular theory in Information Flow (satisfies for example, identity, weakening and global cut). A theory morphism is a (type) function maps source sequents into the target closure. When the target theory is closed, a theory morphism maps source sequents to target sequents. A classification satisfies a sequent when any instance classified by all types in the antecedent is classified by some type in the consequent. The logical environment **IFC** is a subenvironment of the first order logical environment **FOL** when types are regarded as unary relation symbols.

## 5. COALESCENCE

Based upon the horizontal composition of passages and bridges, there is a composition called *coalescence* (Figure 2) from the context product of **V**-diagrams and **V**-indexed contexts to the context of indexed contexts. There is also a dual version of coalescence from on the context product of $\mathbf{V}^{op}$-codiagrams and dual $\mathbf{V}^{op}$-indexed contexts to the context of dual indexed contexts. By using the involution on indexed contexts and the involution on diagrams, these two versions of coalescence can be defined in terms one another: the dual coalescence followed by the indexed context involution is the same as the passage product of the involutions for **V**-diagrams and **V**-indexed contexts followed by coalescence.

However, the dual version of coalescence can be defined directly in terms of horizontal composition. For any ambient context **V**, in the context of codiagrams an object is essentially a passage and a morphism contains a bridge, both with destination context **V**. Also, in the context of dual indexed contexts an object is essentially a passage and a morphism is essentially a bridge, both with originating context **V**. Hence, horizontal composition can be applied to both. The coalescence of a **V**-diagram and a dual **V**-indexed context is a dual indexed context, whose indexing context is that of the diagram and whose passage is the composition of component passages. The coalescence of a **V**-diagram colink and a dual **V**-indexed link is a dual indexed link, whose indexing passage is that of the **V**-diagram colink and whose bridge is the horizontal composition of component bridges.

## 6. FUSION

### 6.1. General Theory

Following the paper (Goguen, 2006), *fusion* [‡] (Figure 2) or the Grothendieck construction (Grothendieck, 1963) is a way for homogeneously handling situations of structural heterogeneity. Such situations are represented by indexed contexts, where one kind of structure is indexed by another, are a central structure found in the institutional approach to ontologies. The fusion process transforms by structural summation an indexed context into a single all-encompassing context. There are two zones involved in fusion: the zone of indexing and the zone of structural summation. The zone of indexing often contains pseudo notions, where structure is unique only up to isomorphism. However, the zone of structural summation is strict. Links in the fused context can represent sharing and translation between objects in the structurally heterogeneous indexed context, and indexed objects in the fused context can be combined using the sum construction. Examples include ontologies.

In overview, the component contexts of an indexed context are assembled together by fusion into a single homogeneous context obtained by forming the disjoint union of their object collections and then adding links based on the indexing links. In detail, given an indexed context, define the fusion context as follows: objects are pairs called indexed objects, which consist of an indexing object and an object in the corresponding component context; and links from one indexed object to another indexed object are pairs called indexed links, which consist of an indexing link and a component link from the object of the beginning indexed object to the contravariantly mapped image of the object of the ending indexed object. Composition of indexed links is defined in terms of composition of the underlying indexing links and component links. The fusion of a dual indexed context is the opposite of the fusion of the corresponding indexed context under index context involution. For any indexed context, the fusion of the indexed context involute has the same indexed objects, but has indexed links whose component link is flipped. The fusion of an inversely indexed context is the fusion of the left component indexed context involute or, by the definition of invertibility, the fusion of the right component dual indexed context. An (dual, inversely) indexed context has a base projection passage from its fusion context to its indexing context. The fibers of an indexed context are the component contexts, whereas the fibers of a dual (inversely) indexed context are the opposite of the component contexts$^§$. The fusion context of a cocomplete dual or inversely indexed context is cocomplete, and its base projection passage is cocontinuous.

The fusion of an indexed link is a passage from the fusion of the originating indexed context to the fusion of the destination indexed context. It maps an indexed object by applying the indexing passage to its indexing object and applying the bridge to its component object, and it maps an indexed link by applying the indexing passage to its indexing link and applying the bridge to its component link. The fusion of an indexed link commutes with its indexing passage through the base projection passages of the origin and destination indexed contexts. The fusion of an indexed connection from one indexed link to another indexed link is a bridge from the fusion passage of the beginning indexed link to the fusion passage of the ending indexed link. The indexing link of the components of the fusion bridge is obtained by applying the bridge of the indexed connection to the indexing objects in the fusion of the beginning indexed context. Hence, the basic fusion process is a 2-dimensional passage from the context of (dual) indexed contexts to the context of contexts. The ordinary and dual versions are definable in terms of one another via the involutions for contexts and indexed contexts. The fusion passage of a cocontinuous dual indexed link is cocontinuous.

For any ambient context **V**, the derived fusion process is the composition of coalescence with basic fusion. As a result, it maps a **V**-diagram and an (dual) **V**-indexed context to the context that is the fusion of their coalescence. It maps a **V**-diagram (co)link and an (dual) **V**-indexed link to the passage that is the fusion of their coalescence. There is an adjoint derived fusion process that maps a **V**-diagram to a process that maps a (dual) **V**-indexed context to the fusion context. The same process applies to links. Here, the (dual) **V**-indexed contexts and links are usually restricted to a relevant subcollection called the core.

### 6.2. *Special Theory*

In the institutional approach, unpopulated ontologies are represented by theories, and populated ontologies are represented by (local) logics. Structures provide an interpretative semantics for languages, theories provide a formal or axiomatic semantics for languages, and logics provide a combined semantics, both interpretative and axiomatic. Theories are the most direct representation for ontologies. The notion of a (local) logic in an institution generalizes the notion of a (local) logic in the theory of Information Flow. Hence, for institutions and institution morphisms the relevant core indexed contexts are structure, inverse flow and logic. The fusions of these comprises the core diagram (Figure 6a) consisting of the context of structures representing interpretative semantics, the context of theories representing formal or axiomatic semantics and the context of logics representing combined semantics, respectively. Since the identity passage on classifications can be regarded as a terminal institution, the core diagram for classifications is just the very special case of the core diagram for this identity institution. The core diagram for an institution is linked to the core diagram for classifications by the classification passage and others built upon it. Each institution morphism generates a link between the core diagrams at origin and destination, which consists of passages for structures, theories and logics. For institutions and institution comorphisms there is only one relevant indexed context: direct flow. The fusion of direct flow is the same theory context given by inverse flow, since inverse and direct flow are components of the same inversely indexed context.

In the institutional approach to ontologies the "lattice of theories" construction is represented as the theory passage from the context of (co)institutions to the context of contexts. For any fixed institution the "lattice of theories" construction is represented "in the large" by the context of theories, and for any language of that institution the "lattice of theories" construction is represented "in the small" by either the concept preorder of theories or the concept lattice of closed theories. From each theory in the order of theories, the entailment order defines paths to the more generalized theories above and the more specialized theories below. There are four ways for moving along paths from one theory to another (Sowa, 2000): contraction, expansion, revision and analogy. Any theory can be contracted or reduced to a smaller, simpler theory by deleting one or more axioms. Any theory can be expanded by adding one or more axioms. A revision step is composite — it uses a contraction step to discard irrelevant details, followed by an expansion step to added new axioms. Unlike contraction, expansion or revision, which move to nearby theories in the order, analogy jumps along a language link to a remote theory in the context (in a first order logic institution this might occur by systematically renaming the entity types, relation types, and constants that appear in the axioms). By repeated contraction, expansion and analogy, any theory can be converted into any other. Multiple contractions would reduce a theory to the empty theory at the top of the lattice. The top theory in the lattice of theories is the closure of the empty theory — it contains only tautologies or logical truths; i.e., sentences that are true in all structures (it is "true of everything"). Multiple expansions would reduce a theory to the full inconsistent theory at the bottom of the lattice. The full inconsistent theory is the closed theory consisting of all expressions; i.e., expressions that are true in no structures (it is "true of nothing").

The context of theories, which is the fusion of the inverse-direct flow, is cocomplete, since inverse-direct flow is a cocomplete inversely **Cls**-indexed order (Tarlecki, Burstall and Goguen, 1991). An institution is said to be cocomplete when its direct flow coalescence, a dual indexed context, is cocomplete; equivalently, when its context of languages is cocomplete. For any institution, the projection passage reflects sums. Hence, if the institution is cocomplete, then its context of (closed) theories is cocomplete and its projection passage is cocontinuous (Goguen and Burstall, 1992). When ontologies are represented by theories (formal or axiomatic representation) in a cocomplete institution, semantic integration of ontologies can be defined via the two steps of alignment and unification[**] (Kent, 2004). When ontologies are represented by (local) logics (interpretative and axiomatic representation) in a logical environment, semantic integration of ontologies can be defined by an analogous process (see the lifting in Figure 6a). An institution colink is cocontinuous when its direct flow coalescence, a dual indexed link, is cocontinuous; equivalently, when it links cocomplete institutions and its language passage is cocontinuous. If an institution colink is cocontinuous, then its theory passage is cocontinuous. For any institution duality (institution colink and link) based on an invertible passage of languages, there is a (closed) invertible passage of theories, with right (left) component being the fusion of the institution (co)link. Hence, the left (closed) theory passage is cocontinuous.

According to the dictionary, a cosmos is an orderly harmonious systematic universe. A polycosmos (Patrick Cassidy) is an unpopulated modular object-level "ontology that has a provision for alternative possible worlds, and includes some alternative logically contradictory theories as applying to alternative possible worlds". The mathematical formulation of polycosmic is given in terms of the sum of a diagram of theories for some institution. A diagram of theories is monocosmic when its sum is consistent (satisfiable by some structure). A diagram of theories is pointwise consistent when each indexed theory in the direct flow along the summing corelation is consistent. A monocosmic diagram of theories is pointwise consistent by default. A diagram of theories is polycosmic when it is pointwise consistent, but not monocosmic; that is, when there are (at least) two consistent but mutually inconsistent theories in the direct flow. In some institutions, there are some extreme polycosmic diagrams of theories, where any two theories are either entailment equivalent (isomorphic) or mutually inconsistent. Each of the theories in these diagrams lies at the lowest level in the lattice of theories, strictly above the bottom inconsistent theory containing all sentences.

7.     FORMALISM

The Information Flow Framework (IFF) (Kent, Farrugia, Schorlemmer and Obrst, 2001–2007) is a descriptive category metatheory under active development, which was originally offered as the structural aspect of the Standard Upper Ontology (SUO) and is now offered as a framework for all theories. The IFF architecture is a two dimensional structure consisting of metalevels (the vertical dimension) and namespaces (the horizontal dimension). Within each level, the terminology is partitioned into namespaces. In addition, within each level, various namespaces are collected together into meaningful composites called meta-ontologies. The IFF is vertically partitioned into the object level at the bottom, the supra-natural part or metashell at the top, and the vast intermediate natural part. The natural part is further divided horizontally into pure and applied aspects. The pure aspect of the IFF is largely concerned with category-theoretic matters. The applied aspect of the IFF is largely governed by the institutional approach to the theory and application of ontologies.

The IFF has had two major developmental phases: experiment and implementation. The experimental phase of the IFF development occurred during the years 2001–2005. The present and future development is mainly concerned with the final coding and the implementation of the IFF. Initially, the plan of development was for the IFF to use category theory to represent various aspects of knowledge engineering, but more recently this strategy was augmented and reversed, thus applying knowledge engineering to the representation of category theory. The institutional approach is the main instrument used by the IFF to connect and integrate axiomatizations of various aspects of knowledge engineering. It is being axiomatized in the upper metalevels of the IFF, and the lower metalevel of the IFF has axiomatized various institutions in which the semantic integration of ontologies has a natural expression as the sum of theories.

Both semantics and formalisms are important for ontologies. The connection between semantics and formalism is through interpretation. The institutional approach is centered on interpretation and represents it as a parameterized relation of satisfaction between semantics and formalism. Although in many common examples the formal side of the satisfaction relation is set-theoretically small and the semantical side is set-theoretically large, in the IFF axiomatization of the institutional approach both sides can range through the hierarchy of metalevels. Hence, we think of the institutional approach as existing at a higher level, with its domain including the triad (semantics, formalism, interpretation) and its formal system encoded in category-theoretic terms. However, it incorporates category theory also, with the contents of category theory as its semantics and the axiomatization of category theory (as done in the IFF) as its formal system. So category theory provides a formalism for the institutional approach, and the institutional approach provides an interpretation for category theory.

---

[*] Here is a key to the terminology used in this paper.

| this paper | category theory |
|---|---|
| object | object |
| link | arrow, morphism, 1-cell |
| connection | 2-cell |
| beginning, origin(ation) | source, domain |
| ending, destination | target, codomain |
| context | category |
| passage | functor |
| bridge | natural transformation |
| invertible passage | adjunction |
| equivalence | natural equivalence |
| (co)relation | (co)cone |
| sum (product) | colimit (limit) |
| relative sum | left Kan extension |

† The category **Cls** of classifications and infomorphisms is the ambient category that is used for indexing in the institutional approach to ontologies. This is the category of "twisted relations" of (Goguen and Burstall, 1992). This is also the basic category used in the theory of Information Flow and Formal Concept Analysis (Kent, 2002).

‡ The fusion of an indexed context might be called "fusion in the large" or "structural fusion". The sums of diagrams, in particular sums of diagrams of theories of an institution, which takes place within the fused context of theories, might be called "fusion in the small" or "theoretical fusion". Both are kinds of constrained sums.

§ This explains the flip in the definition of cocompleteness and cocontinuity for dual (inversely) indexed contexts.

∗∗ We describe the semantic integration of ontologies in terms of theories.

**Alignment:** Informally, identify the theories to be used in the construction. Decide on the semantic interconnection (semantic mapping) between theories. This may involve the introduction of some additional mediating (reference) theories. Formally, create a diagram of theories of shape (indexing) context that indicates this selection and interconnection. This diagram of theories is transient, since it will be used only for this computation. Other diagrams could be used for other sum constructions. Compute the base diagram of languages with the same shape. Form the sum language of this diagram, with language summing corelation. Being the basis for theory sums, language sums are important. They involve the two opposed processes of "summing" and "quotienting". Summing can be characterized as "keeping things apart" and "preserving distinctness", whereas quotienting can be characterized as "putting things together", "identification" and "synonymy". The "things" involved here are symbolic, and for a first order logic institution may involve relation type symbols, entity type symbols and the concepts that they denote.

**Closure:** Form the sum theory of the diagram of theories, with theory summing corelation. The summing corelation is a universal corelation that connects the individual theories in the diagram to the sum theory. The sum theory may be virtual. Using direct flow, move the individual theories in the diagram of theories from the "lattice of theory diagrams over the language diagram" along the language morphisms in the language summing corelation to the lattice of theories over the language sum, getting a homogeneous diagram of theories with the same shape, where each theory in this direct flow image diagram has the same sum language (the meaning of homogeneous). Compute the meet of this direct flow diagram within the fiber "lattice of theories" over language sum, getting the sum theory. The language summing corelation is the base of the theory summing corelation. Using inverse flow, move the sum theory from the language sum back along the language morphisms in the language summing corelation to the language diagram, getting the system closure.